\documentclass[11pt]{article}
\usepackage{amsmath,amssymb,color,graphics,epsfig,cite}

\usepackage{amsfonts}

\newcommand{\hoch}[1]{$\, ^{#1}$}

\makeatletter
\@addtoreset{equation}{section}
\makeatother

\newcommand{\be}{\begin{equation}}
\newcommand{\ee}{\end{equation}}
\newcommand{\bea}{\setlength\arraycolsep{2pt} \begin{eqnarray}}
\newcommand{\eea}{\end{eqnarray}}
\newcommand{\nn}{\nonumber}

\def\ft#1#2{{\textstyle{\frac{\scriptstyle #1}{\scriptstyle #2} } }}
\def\fft#1#2{{\frac{#1}{#2}}}

\def\0{{\sst{(0)}}}
\def\1{{\sst{(1)}}}
\def\2{{\sst{(2)}}}
\def\3{{\sst{(3)}}}
\def\4{{\sst{(4)}}}
\def\5{{\sst{(5)}}}
\def\6{{\sst{(6)}}}
\def\7{{\sst{(7)}}}
\def\8{{\sst{(8)}}}
\def\sst#1{{\scriptscriptstyle #1}}

\begin{document}


\begin{center}
{\Large {\bf Higher-Derivative Gravity with Non-minimally Coupled Maxwell Field}}

\vspace{40pt}
{\bf Xing-Hui Feng and H. L\"u}

\vspace{10pt}

\hoch{1}{\it Center for Advanced Quantum Studies, Department of Physics, \\
Beijing Normal University, Beijing 100875, China}

\vspace{40pt}

\underline{ABSTRACT}
\end{center}

We construct higher-derivative gravities with a non-minimally coupled Maxwell field.  The Lagrangian consists of polynomial invariants built from the Riemann tensor and the Maxwell field strength in such a way that the equations of motion are second order for both the metric and the Maxwell potential. We also generalize the construction to involve a generic non-minimally coupled $p$-form field strength.  We then focus on one low-lying example in four dimensions and construct the exact magnetically-charged black holes.  We also construct exact electrically-charged $z=2$ Lifshitz black holes.  We obtain approximate dyonic black holes for the small coupling constant or small charges.  We find that the thermodynamics based on the Wald formalism disagrees with that derived from the Euclidean action procedure, suggesting this may be a general situation in higher-derivative gravities with non-minimally coupled form fields. As an application in the AdS/CFT correspondence, we study the entropy/viscosity ratio for the AdS or Lifshitz planar black holes, and find that the exact ratio can be obtained without having to know the details of the solutions, even for this higher-derivative theory.

\vfill {\footnotesize xhfengp@mail.bnu.edu.cn \ \ \ mrhonglu@gmail.com}

\thispagestyle{empty}

\pagebreak

\tableofcontents
\addtocontents{toc}{\protect\setcounter{tocdepth}{2}}



\section{Introduction}

Spacetime metric $g_{\mu\nu}$, the nonlinear generalization of the massless spin-2 field, is the fundamental field in the Einstein's formulation of gravity. Electric-magnetic interactions of the $U(1)$ Maxwell field $A_\mu$ underlies almost all the phenomena in condensed matter physics.  With the development of the AdS/CFT correspondence \cite{mald,gkp,wit}, the Einstein-Maxwell theory with a negative cosmological constant becomes one of the most important playgrounds in relating classical gravity to certain strongly-coupled condensed matter theories (CMT) at the quantum level, from superconductivity \cite{Hartnoll:2008vx} to Non-Fermi liquids \cite{Lee:2008xf,lmv}.

Whilst there have been great progresses in studying condensed matter physics via gravity, the successes are mainly in qualitative nature.  To match a condensed matter phenomenon quantitatively as well, it is likely that one needs to generalize the Einstein-Maxwell theory, by introducing additional fields and/or couplings.  One generalization, without breaking the general coordinate invariance, is to consider higher-derivative extensions.  Higher-derivative gravity arises naturally in string or M-theory, where the AdS/CFT correspondence has the most solid foundation. The low-energy effective theories of string or M-theory are supergravities as the leading-order expansions, with some specific but infinite sequences of higher-derivative corrections. Einstein-Maxwell gravities in both four and five dimensions can be supersymmetrized, and embedded in M-theory \cite{Pope:1985jg,Pope:1985bu} or the type IIB string \cite{Chamblin:1999tk,Cvetic:1999xp}. (A specific extra $FFA$ term is necessary in $D=5$ for supersymmetrization.) It is thus natural to consider higher-derivative extensions for the Einstein-Maxwell theory.

When a linear theory involves higher derivative terms, there are inevitable ghost excitations.  This problem can be easily circumvented via nonlinear construction for scalar, vector and anti-symmetric tensor fields.  This is because for these fields, the first derivative is also a tensor or can be made a tensor, without breaking the gauge symmetries.  One can then construct a higher-derivative theory by adding higher-order polynomial invariants of these fields and/or their tensorial first derivatives.  Although the theory may involve high-order total derivatives through nonlinearity, each field has at most two derivatives acting upon directly in the equations of motion.  Consequently, the linearized theory in any background is of the second order.  The situation is rather different for the metric.  The first derivative of the metric cannot be a non-vanishing tensor and only two derivatives of the metric may yield a tensor, namely the Riemann tensor.  It follows that a typical higher-order polynomial invariant of the Riemann tensor tends to give rise to linear ghost excitations in a generic background.

There are different approaches concerning the ghost issue in higher-derivative gravities.  In supersymmetric theories, ghosts may not be fatal \cite{Smilga:2013vba}.  In fact in four dimensions, gravity extended with quadratic curvature invariants were shown to be renormalizabe \cite{Stelle:1976gc,Stelle:1977ry}.  Recently a new static black hole over and above the usual Schwarzschild black hole were obtained in the four-dimensional theory \cite{Lu:2015cqa}.  When there is a cosmological constant, higher-derivative gravities in AdS backgrounds can have a critical point in the parameter space for which the ghost modes become log modes and may be truncated out by some strong boundary conditions.  However, this process was more successful in three dimensions \cite{Li:2008dq,Bergshoeff:2009hq} than in four or higher dimensions \cite{Lu:2011zk,Deser:2011xc,Porrati:2011ku}.

In perturbative string theory, the coupling constants of higher-order terms are regarded as small.  One may use the field redefinition of the metric
\be
g_{\mu\nu} \rightarrow g_{\mu\nu} + \alpha R g_{\mu\nu} + \beta R_{\mu\nu}\,,\label{gmunuredef}
\ee
to simplify the theory order by order. In this approach, the propagators are not modified and hence the ghost issue does not arise, even though the theory would have ghosts when treated own its own.  The shortcoming is that the contributions from the higher-order terms can only be regarded as small.  This is too restrictive in the applications of the the AdS/CFT correspondence, since in the discussion of gravity/CMT the purpose of introducing higher-order terms is not simply to add small perturbation.

It turns out that there are combinations of polynomial invariants that are ghost free.  The most famous example is the Gauss-Bonnet term.  Einstein gravity extended with the Gauss-Bonnet term has a total of four derivatives via nonlinearity, but is ghost free since the theory involves only two derivatives at the linear level.  Consequently the coupling constant of the Gauss-Bonnet term does not have to be small.  (Causality consideration may provide further restrictions on the coupling constant \cite{Hofman:2008ar,deBoer:2009pn,Camanho:2009vw,Buchel:2009sk}.) The Gauss-Bonnet term is one of a class of Euler integrands that give rise to general Lovelock gravities \cite{Lovelock:1971yv}.  These theories are making sense only in the context of string theory.  Firstly, Gauss-Bonnet gravity violates causality on general grounds and the only way to avoid this problem is by adding an infinite tower of massive higher-spin particles \cite{Camanho:2014apa}.  Secondly, $D=10$, ${\cal N}=1$ supergravity with the string worldsheet $\alpha'$ correction indeed have a Riemann-squared \cite{Bergshoeff:1989de}
\be
\alpha' R^{\mu\nu\rho\sigma}R_{\mu\nu\rho\sigma}\nn
\ee
correction.  Using the field redefinition (\ref{gmunuredef}), one can generate the Gauss-Bonnet term at the quadratic order of the curvature polynomials.  In other words, the Gauss-Bonnet term or higher-order Euler integrands arise naturally in string theory.  One may then appeal to the enormity of the string landscape and argue that in some string vacua, the Gauss-Bonnet term dominates and hence the Einstein-Gauss-Bonnet gravity can be treated on its own.

In this paper, we generalize this line of approach to include the Maxwell field, or more general $p$-form field strengths as well.  We construct general higher-derivative gravities coupled to Maxwell field with the Lagrangian built from polynomial invariants of the Riemann tensor and the Maxwell field strength.  We require that in all the equations of motion both the metric $g_{\mu\nu}$ and $A_\mu$ have at most two derivatives acting on directly so that the theory may be ghost free.  Since the field strength couples to the curvature tensor directly, the Maxwell field is non-minimally coupled, and also the gauge symmetry is preserved.  Such couplings arise naturally in string theory and we expect that through field redefinition analogous to (\ref{gmunuredef}), ghost-free combinations can also emerge, as in the case of Einstein-Gauss-Bonnet gravity or more general Lovelock gravities.

We now give the outline of the paper.  In section 2, we construct higher-derivative gravities whose Lagrangian consists of the polynomial invariants of Riemann tensor and the field strength $F_{\mu\nu}=\partial_\mu A_\nu - \partial_\nu A_\mu$.  Analogous to the Euler integrands in Lovelock gravities, the combination of the polynomials is such that the equations of motion are second order.  In section 3, we generalize the construction to involve a generic non-minimally coupled $p$-form field strength.  In section 4, we consider a low-lying example in which the Einstein-Maxwell theory with a cosmological constant is augmented with the polynomial of the Riemann tensor with a bilinear of $F_{\mu\nu}$ so that the theory has at most four total derivatives.  The equations of motion nevertheless remain second order.  We construct static charged black holes in four dimensions with isometries of 2-sphere, 2-torus and hyperbolic 2-space.  In section 5, we study an application of the AdS/CFT correspondence and derive the boundary viscosity/entropy ratio for the AdS and Lifshitz planar black holes.  We conclude the paper in section 6.

\section{Non-minimally coupled Maxwell field}

\subsection{The general construction}

Our construction is analogous to Lovelock gravities, whose basic ingredients are Euler integrands, defined by
\be
E^{(k)}=\fft{1}{2^{k}} \delta_{a_1b_1\cdots a_k b_k}^{c_1d_1\cdots c_kd_k}
R^{a_1b_1}_{c_1d_1}\cdots R^{a_kb_k}_{c_kd_k}\,,
\ee
where $R^{ab}_{cd}$ denotes the Riemann tensor $R^{ab}{}_{cd}$ and
\be
\delta_{\alpha_1\cdots\alpha_{s}}^{\beta_1\cdots\beta_{s}}=
s! \delta_{[\alpha_1}^{\beta_1} \cdots \delta_{\alpha_s]}^{\beta_s}\,.
\ee
The Euler integrands can also be expressed as
\be
E^{(k)} = \ft{(2k)!}{2^k}\, R^{[a_1b_1}_{a_1b_1} \cdots R^{a_k b_k]}_{a_k b_k}\,.
\ee
The low-lying examples are
\be
E^{(0)} =1\,,\qquad
E^{(1)}=R\,,\qquad E^{(2)} = R^2 - 4R^{\mu\nu}R_{\mu\nu} + R^{\mu\nu\rho\sigma}R_{\mu\nu\rho\sigma}\,,\quad etc.
\ee
The term $\sqrt{-g} E^{(k)}$ in the Lagrangian contributes
\be
E_{\mu}^{(k)\,\nu} = -\ft{1}{2^{k+1}} \delta_{c_1d_1\cdots c_k d_k\,\mu}^{a_1b_1\cdots a_k b_k\,\nu}\,
R^{a_1b_1}_{c_1d_1}\cdots R^{a_k b_k}_{c_k d_k}
\ee
to the Einstein's equation of motion.  A striking property is that no Riemann-tensor factor acquires any derivative in the equations of motion, such that the theory remains second order in derivatives.  This is a consequence of the fact that the variation of the Riemann tensor, namely
\be
\delta R^{\mu}{}_{\nu\rho\sigma} = \nabla_\rho \Gamma^{\mu}_{\sigma \nu} -
 \nabla_\sigma \Gamma^{\mu}_{\rho \nu}
\ee
yields a total derivative in the Lagrangian for the polynomial combinations of the Euler integrands.  This is largely due to the Bianchi identity of the Riemann tensor, namely
\be
\nabla_{[\alpha} R^{\mu\nu}_{\rho\sigma]}=0=\nabla^{[\beta} R^{\mu\nu]}_{\rho\sigma}\,.\label{rprop}
\ee
In order to include Maxwell field $A$ in an analogous construction, we introduce a bilinear tensor of the field strength $F=dA$
\be
Z^{ab}_{cd} = F^{ab}F_{cd}\,.
\ee
This tensor shares some similar properties of the Riemann tensor, but the properties (\ref{rprop}) and $R^a{}_{[bcd]}=0$ of the Riemann tensor do not extend to the $Z$ tensor.  Nevertheless, owing to the Bianchi identity of the Maxwell field, namely
\be
\nabla_{[\alpha} F_{\rho\sigma]}=0=\nabla^{[\beta} F^{\mu\nu]}\,,
\ee
the $Z$ tensor satisfies the property
\be
\nabla_{[\alpha} \nabla^{[\beta} Z^{\mu\nu]}_{\rho\sigma]}=\nabla_{[\alpha}
F^{[\mu\nu} \nabla^{\beta]} F_{\rho\sigma]} +
2F^{[\mu\nu}\,R^{\beta]}{}_{[\rho\sigma}{}^\lambda\, F_{\alpha]\lambda}\,.
\label{maxprop}
\ee
In other words, although each term involves a total of four derivatives, both $A_\mu$ and $g_{\mu\nu}$ have at most two derivatives.  This property is crucial in our construction.

With these preliminaries, we consider polynomial invariants of the tensor $R^{ab}_{cd}$ and $Z^{ab}_{cd}$ analogous to the Euler integrands, namely
\bea
L^{(m,n)} &=&\fft{1}{2^{m+n}} \delta_{a_1b_1\cdots a_m b_m \tilde a_1\tilde b_1\cdots \tilde a_n \tilde b_n}^{c_1d_1\cdots c_m d_m \tilde c_1\tilde d_1\cdots \tilde c_n \tilde d_n} R^{a_1b_1}_{c_1d_1}\cdots R^{a_m b_m}_{c_m d_m}
Z^{\tilde a_1\tilde b_1}_{\tilde c_1\tilde d_1} \cdots Z^{\tilde a_m\tilde b_m}_{\tilde c_m\tilde d_m}\cr
&=& \ft{(2(m+n))!}{2^{m+n}}\,
R^{[a_1b_1}_{a_1b_1}\cdots R^{a_m b_m}_{a_m b_m}
Z^{\tilde a_1\tilde b_1}_{\tilde a_1\tilde b_1} \cdots Z^{\tilde a_m\tilde b_m]}_{\tilde a_m\tilde b_m}\,.
\eea
It is clear that when $n=0$, the above gives rise to the Euler integrands, i.e.
\be
L^{(k,0)}=E^{(k)}\,.
\ee
It is easy to perform the variation of both the metric and $A$:
\be
\delta \big(\sqrt{-g} L^{(m,n)}\big) = \sqrt{-g}\Big(L^{(m,n)}_{(\mu\nu)} \delta g^{\mu\nu} + L^{(m,n)\,\mu} \delta A_\mu\Big)+ \hbox{total derivatives}\,.
\ee
We find
\bea
L^{(m,n)}_{\mu\nu}&=&-\ft12 g_{\mu\nu}L^{(m,n)}\nn\\
&&+\frac{2n}{2^{m+n}}\delta^{c_1d_1\cdots c_md_m\tilde c_1\tilde d_1\cdots\tilde c_n\tilde d_n}_{a_1b_1\cdots a_mb_m\tilde a_1\mu\cdots\tilde a_n\tilde b_n}R^{a_1b_1}_{c_1d_1}\cdots R^{a_mb_m}_{c_md_m}Z^{\tilde a_1}_{~~\nu\tilde c_1\tilde d_1}\cdots Z^{\tilde a_n\tilde b_n}_{\tilde c_n\tilde d_n}\nn\\
&&+\frac{m}{2^{m+n}}\delta^{c_1d_1\cdots c_md_m\tilde c_1\tilde d_1\cdots\tilde c_n\tilde d_n}_{a_1\mu\cdots a_mb_m\tilde a_1\tilde b_1\cdots\tilde a_n\tilde b_n}R^{a_1}_{~~\nu c_1d_1}\cdots R^{a_mb_m}_{c_md_m}Z^{\tilde a_1\tilde b_1}_{\tilde c_1\tilde d_1}\cdots Z^{\tilde a_n\tilde b_n}_{\tilde c_n\tilde d_n}\nn\\
&&+\frac{2m}{2^{m+n}}g_{c_1\mu}\delta^{c_1d_1\cdots c_md_m\tilde c_1\tilde d_1\cdots\tilde c_n\tilde d_n}_{\nu b_1\cdots a_mb_m\tilde a_1\tilde b_1\cdots\tilde a_n\tilde b_n}R^{a_2b_2}_{c_2d_2}\cdots R^{a_mb_m}_{c_md_m}\nabla^{b_1}\nabla_{d_1}(Z^{\tilde a_1\tilde b_1}_{\tilde c_1\tilde d_1}\cdots Z^{\tilde a_n\tilde b_n}_{\tilde c_n\tilde d_n}),\nn\\
L^{(m,n)\mu} &=& \nabla_\nu{\widehat F}^{\mu\nu}\,,\cr
{\widehat F}^{\mu\nu}&=&\frac{4n}{2^{m+n}}\delta^{c_1d_1\cdots c_md_m\mu\nu\cdots\tilde c_n\tilde d_n}_{a_1b_1\cdots a_mb_m\tilde a_1\tilde b_1\cdots\tilde a_n\tilde b_n}R^{a_1b_1}_{c_1d_1}\cdots R^{a_mb_m}_{c_md_m}F^{\tilde a_1\tilde b_1}Z^{\tilde a_2\tilde b_2}_{\tilde c_2\tilde d_2}\cdots Z^{\tilde a_n\tilde b_n}_{\tilde c_n\tilde d_n}\,.
\eea
It follows from (\ref{rprop}) and (\ref{maxprop}) that neither the metric nor $A$ has more than two derivatives in all terms in $L^{(m,n)}_{(\mu\nu)}$ and $L^{(m,n)\,\mu}$.

The Lagrangian for the general theory is then given by
\be
{\cal L}=\sqrt{-g} \sum_{k=0}\,\sum_{m+n=k} \gamma_{mn} L^{(m,n)}\,,
\ee
where $\gamma_{mn}$ are coupling constants.
The full set of equations of motion are
\be
\sum_{k}\, \sum_{m+n=k} \gamma_{mn} L^{(m,n)}_{(\mu\nu)}=0\,,\qquad
\sum_{k}\, \sum_{m+n=k} \gamma_{mn} L^{(m,n)\,\mu}=0\,.
\ee
Again, in all these equations, the metric and $A_\mu$ have at most two derivatives acting on directly, with the total higher derivatives achieved through nonlinearity.  The theories are thus of the second order.

We note that the non-minimally coupled Maxwell field can also have the following structure
\be
\sqrt{-g}\, R^{[a_1b_1}_{a_1b_1}\cdots R^{a_n b_n}_{a_n b_n} F^{c_1}{}_{c_1}\cdots
F^{c_k]}{}_{c_k}\,.\label{str2}
\ee
When $k=1$, this term is a total derivative.  When $k=2n+1$ with $n\ge 1$, this term vanishes.  For $k=2n$, this term is proportional to the $\sqrt{-g} L^{(m,n)}$ owing to the identity $R_{[abc]d}=0$.  Thus we shall not consider the terms (\ref{str2}).  It is also worth pointing out again that any polynomial structures involving purely the Maxwell field strength without the Riemann tensor are allowed and hence we shall not list them all.

\subsection{A low-lying example}

Having constructed general higher-derivative gravities with non-minimally coupled Maxwell field, we shall study a low-lying example in detail. The Lagrangian is
\be
{\cal L}=\sqrt{-g}\Big(R-2\Lambda_0 - \ft14 F^2 + \gamma L^{(1,1)}\Big)\,,\label{focus}
\ee
where
\be
L^{(1,1)} = \ft14 \delta^{cd\tilde c\tilde d}_{ab\tilde a\tilde b}
R^{ab}_{cd} Z^{\tilde a\tilde b}_{\tilde c\tilde d}= RF^2 - 4 R_{ab} F^{ac}F^b{}_c + R_{abcd} F^{ab} F^{cd}\,.
\ee
In other words, the theory is the Einstein-Maxwell theory with a cosmological constant, together with an additional $L^{(1,1)}$ term. The Einstein equations of motion are
\be
G_{\mu\nu} + \Lambda_0 g_{\mu\nu} - \ft12 (F_{\mu\nu}^2 -\ft14 g_{\mu\nu} F^2) +
\gamma L^{(1,1)}_{(\mu\nu)}=0\,,\label{eomabs}
\ee
where
\bea
L^{(1,1)}_{\mu\nu} &=& -\ft12 g_{\mu\nu} L^{(1,1)} +
\ft12 \delta^{cd\tilde c\tilde d}_{ab\tilde a\mu}\, R^{ab}_{cd}\, F_{\tilde c\tilde d} F^{\tilde a}{}_\nu +\ft14 \delta^{cd\tilde c\tilde d}_{a\mu\tilde a\tilde b}\,
R^a{}_{\nu cd}\, Z^{\tilde a\tilde b}_{\tilde c\tilde d}\cr
&&+\ft12 g_{c\mu} \delta^{cd\tilde c \tilde d}_{\nu b \tilde a \tilde b}\,
 \nabla^b \nabla_d\,(Z^{\tilde a\tilde b}_{\tilde c\tilde d})\,.
\eea
The Maxwell equation is
\be
\nabla_{\mu} \widehat F^{\mu\nu}=0\,,\quad\hbox{with}\qquad
\widehat F^{\mu\nu}\equiv F^{\mu\nu} - \delta^{cd\mu\nu}_{ab\tilde a\tilde b} R^{ab}_{cd} F^{\tilde a\tilde b}\,.\label{maxabs}
\ee
Owing to the Bianchi identity of the Riemann tensor, the differential operator $\nabla_\mu$ can only land on $F$, but not $R$, and hence the theory is of the second order.  In section 4, we shall construct charged black holes of this theory.

\section{Non-minimally coupled $p$-form field strength}

The construction in the previous section can be easily generalized to general $(p-1)$-form potential $A_{(p-1)}$ whose $p$-form field strength is given by
\be
F_{(p)}=dA_{(p-1)}\,,\qquad F_{a^1\cdots a^p}=p\, \nabla_{[a^1} A_{a^2\cdots a^p]}\,.
\ee
For the simplicity of notations, we construct the corresponding $Z$ tensors
\be
Z^{a^1\cdots a^p}_{b^1\cdots b^p} =
F^{a^1\cdots a^p}F_{b^1\cdots b^p}\,.
\ee
The generalizing polynomial of the $p$-form to $L^{(m,n)}$ of the 2-form field strength is then given by
\be
L^{(m,n),p}=\fft{(2m+pn)!}{2^m (p!)^n} R^{[a_1b_1}_{a_1b_1}\cdots R^{a_m b_m}_{a_m b_m} Z^{a^1_1\cdots a^p_1}_{a^1_1
\cdots a^p_1} \cdots Z^{a^1_n\cdots a^p_n]}_{a^1_n
\cdots a^p_n}\,,\label{generalp}
\ee
Owing to the Bianchi identity
\be
\nabla^{[a^{p+1}} F^{a^1\cdots a^p]}=0=\nabla_{[b^{p+1}} F_{b^1\cdots b^p]}\,,
\ee
it is straightforward to verify that in the equations of motion associated with the Lagrangian
\be
\sqrt{-g} L^{(m,n),p}\,,\nn
\ee
neither the metric nor $A_{(p-1)}$ has more than two derivatives, even though the theory involves higher-order derivatives through nonlinearity.  When $p$ is odd, we have $L^{(m,n),p}=0$ for $n\ge 2$.  Note that for $p=1$, we must have $n=0, 1$. The series $L^{(m,1),1}\equiv H^{(m)}$ was first constructed by Horndeski \cite{Horndeski:1974wa}. The $p=2$ series was constructed in the previous section.

It should be pointed out that the non-minimal coupling terms $L^{(m,n),p}$ are not the only possible structures that one can build for ghost-free combinations. For example, when $p=3$, we can also have terms like
\be
\fft{(2m+3n)!}{2^m 6^n} R^{[a_1b_1}_{a_1b_1}\cdots R^{a_m b_m}_{a_m b_m} Y^{a^1_1\cdots a^3_1}_{a^1_1
\cdots a^3_1} \cdots Y^{a^1_n\cdots a^3_n]}_{a^1_n
\cdots a^3_n}\,,\label{p=3new}
\ee
with
\be
Y^{a^1 a^2 a^3}_{b^1 b^2 b^3}= F^{a^1 a^2}{}_{b^1} F^{a^3}{}_{b^2 b^3}\,.
\ee
It is fairly straightforward to verify that the equations of motion are second order. The most dangerous term that can arise in the equations of motion is
\be
\nabla^{[a^1}\nabla_{[b^1} Y^{a^2 a^3 a^4]}_{b^2 b^3 b^4]}\,.\label{nablaa1b1}
\ee
It is useful to note that
\be
F^{a b}{}_c=2\nabla^{[a} A^{b]}{}_c + \nabla_c A^{ab}\,.
\ee
It then becomes obvious that (\ref{nablaa1b1}) does not involve three or more derivatives derivatives.

As $p$ increases, more and more possible ghost-free polynomial structures can be built. We shall not in this paper classify all such terms for general $p$-forms.  It is also worth pointing out that in the construction, we can replace the $p$-form field strength with the $p$-form potentials, whose kinetic term needs to be further introduced.  The corresponding theory may also be ghost free.  In particular the Einstein-vector theory was constructed in \cite{Geng:2015kvs}.

\section{Electric and magnetic black holes}

\subsection{Static ansatz and reduced equations of motion}

In this section, we focus on the low-lying four-derivative theory (\ref{focus}) in four dimensions, where the Gauss-Bonnet term is a total derivative and hence irrelevant. We construct static black holes that carry electric and magnetic charges.  The ansatz is given by
\bea
ds^2 &=& -h dt^2 + \fft{dr^2}{f} + r^2 d\Omega_{2,\epsilon}^2\,,\cr
A&=& \phi\, dt + p\, \omega_\1\,,\qquad d\omega_\1 = \Omega_{\2}^\epsilon\,,
\label{ansatz}
\eea
where $p$ is a constant.  The metric functions $(h, f)$ and the electrostatic potential $\phi$ are functions of $r$.  The metric $d\Omega_{n-2,\epsilon}^2$ of the level surfaces is
\be
d\Omega_{2,\epsilon}^2 = \fft{dx^2}{1-\epsilon x^2} +
    (1-\epsilon x^2)\, dy^2\,.\label{d2metric}
\ee
The topology parameter $\epsilon$ takes values of $1,0,-1$, for the unit
$S^{2}$, the $2$-torus or the unit hyperbolic $2$-space.  The 1-form $\omega_\1$ is simply $\omega_\1=x dy$ and $\Omega_{\2}^\epsilon=dx\wedge dy$ is the volume 2-form for the metric (\ref{d2metric}).  With these conventions, we see instantly that the ansatz carries the magnetic charge
\be
Q_m=\fft{1}{16\pi} p \int \Omega_{\2}^\epsilon=\fft{\omega_{2,\epsilon}}{16\pi} p=\ft1{4}p\,.
\ee
Throughout this paper, we set, without loss of generality, the volume $\omega_{2,\epsilon}$ of level surfaces to be independent of the topology, namely
\be
\omega_{2,\epsilon}=4\pi\,, \qquad \hbox{for}\qquad \epsilon=1,0,-1\,.
\ee
For $\epsilon=1$, it is the true volume of the $S^2$. For $\epsilon=0$ the extensive quantities such as mass and charges are then density quantities per $4\pi$ area.

The ansatz (\ref{ansatz}) is the most general one for the static configuration with isometries of either $S^2$, $T^2$ or $H^2$.  The Maxwell equation (\ref{maxabs}) becomes\be
\Big(\big(8\gamma (f-\epsilon) + r^2\big) \sqrt{\ft{f}{h}}\, \phi'\Big)'=0\,.
\ee
The first integral can be easily obtained as a quadrature
\be
\phi'=\fft{q}{8\gamma(f-\epsilon) +r^2}\sqrt{\fft{h}{f}}\,,
\ee
where $q$ is an integration constant.  This determines the electric charge, given by
\be
Q_e=\fft{1}{16\pi} \int \sqrt{-g} \widehat F^{01}=\fft{\omega_{2,\epsilon}}{16\pi} q = \ft14 q\,,
\ee
where $\widehat F$ is defined in (\ref{maxabs}).

The Einstein equations (\ref{eomabs}) can now be reduced to one first-order nonlinear differential equation and one quadrature:
\bea
f' &=&- \fft{1}{4(r^4-2\gamma p^2)}\Big(4r^3 (f -\epsilon + \Lambda_0 r^2) +
\fft{q^2 r^3}{8\gamma(f-\epsilon) + r^2} +
\fft{p^2 (48\gamma f + r^2)}{r}\Big)\,,\cr
h &=& u\, f \qquad \fft{u'}{u} = \fft{4\gamma}{r^4-2\gamma p^2}
\Big(\fft{3p^2}{r} - \fft{q^2 r^3}{8\gamma(f-\epsilon) + r^2}\Big)\,.\label{einseom}
\eea

\subsection{General properties}

Many information can be extracted without solving the equations (\ref{einseom}).  The general solution is expected to be parametrized by three quantities, namely the mass and electric and magnetic charges, $(\ft14 q,\ft14 p)$.  The near-horizon geometry is then specified by the horizon radius $r_0$, for which $f(r_0)=0$, and $(q,p)$.  It follows from (\ref{einseom}) that
\be
h'(r_0) = u(r_0) f'(r_0)\,,\qquad
f'(r_0) =\fft{r_0}{4(r_0^4 - 2\gamma p^2)}
\Big(4r_0^2 (\epsilon-\Lambda_0 r_0^2) -
p^2 - \fft{q^2}{1-\fft{8\epsilon\gamma}{r_0^2}}\Big)\,.
\ee
The temperature of the black hole can then easily determined by the standard technique:
\be
T=\fft{\sqrt{h'(r_0)f'(r_0)}}{4\pi}=\fft{f'(r_0) \sqrt{u(r_0)}}{4\pi}\,.
\ee
Entropy can be obtained using the Wald entropy formula \cite{wald1,wald2}
\be
S=-\ft18\int d^{n-2}x\,\sqrt{-h} \fft{\partial({\cal L}/\sqrt{-g})}{\partial R^{abcd}} \epsilon^{ab} \epsilon^{cd}\,,
\ee
which yields
\be
S=\pi r_0^2 \big(1 + \fft{2\gamma p^2}{r_0^4}\big)=\pi r_0^2 \Big(1 + \fft{32\gamma Q_m^2}{r_0^4}\Big)\,.\label{entropy}
\ee
It is worth commenting that the Wald entropy formula is not always valid.  It was shown to be invalid in Einstein-Horndeski gravity, owing to the unusual behavior of the scalar in black hole horizon \cite{Feng:2015oea,Feng:2015wvb}. In our charged black holes, however, the Maxwell field behaves in the similar fashion as the Reissner-Nordstr\o m (RN) black hole on the horizon and hence we expect that the Wald entropy formula holds in our black hole solutions.

The asymptotic region is less universal.  For generic parameters, the large $r$ expansions for $f$ and $u$ are
\bea
f &=&-\ft13\Lambda_0 r^2 + \epsilon -\fft{\mu}{r} -\Big(\ft1{12}(32\gamma \Lambda_0 -3)p^2 + \fft{3q^2}{4(8\gamma \Lambda_0-3)}\Big)\fft{1}{r^2}+ \cdots\,,\cr
\fft{u}{u_0} &=& 1 - \fft{3\gamma}{r^4} \Big(p^2 - \fft{3q^2}{(8\gamma\Lambda_0 - 3)^2}\Big) + \cdots
\eea
This expansion becomes singular when
\be
8\gamma \Lambda_0=3\,,\qquad \hbox{and}\qquad q\ne 0\,.\label{specialgamma1}
\ee
As we shall see presently that the solution describes the $z=2$ charged Lifshitz black hole for these special parameters (\ref{specialgamma1}).

It is worth commenting that as was shown in \cite{Liu:2015tqa} for purely electric AdS planar black holes ($p=0$ and $\epsilon=0$), there is global scaling symmetry whose conserved Noether charge is given by
\be
Q_N=\ft14 \sqrt{\ft{f}{h}}\,\Big(-2r h + r^2 h' - (r^2 + 8\gamma f) \phi\phi' +
4\gamma r f\phi'^2\Big)\,.
\ee
It is easy to verify that evaluating both on the horizon and asymptotic (A)dS infinity yields
\bea
Q_N\Big|_+ &=& T S\,,\qquad S=\pi r_0^2\,,\cr
Q_N\Big|_\infty &=& \ft34\mu -\ft12\phi_0 q=\ft32M - \Phi_e Q_e\,.
\eea
The conservation of the Noether charge implies the following generalized Smarr relation
\be
M=\ft23 (TS + \Phi_e Q_e)\,.\label{smarr1}
\ee

\subsection{Exact general magnetic black holes}

When $q=0$, the ansatz (\ref{ansatz}) carries only magnetic charges.  In this case, the equations can be solved completely, given by
\be
u=\Big(1 - \fft{2\gamma p^2}{r^4}\Big)^{\fft32}\,,
\ee
and
\bea
f &=& u^{-\fft76}\Big( \fft{(3p^2 + 48\epsilon r^2 - 16\Lambda_0 r^4)\sqrt{u}}{48r^2} - \fft{\mu}{r}\cr
&&\fft{(3-32\gamma\Lambda_0)p^2}{16r^2}{}_2F_1[\ft14,\ft14;\ft54;
\ft{2\gamma p^2}{r^4}] + \fft{2\epsilon\gamma p^2}{r^4}
{}_2F_1[\ft14,\ft34;\ft74;\ft{2\gamma p^2}{r^4}]\Big)\,.
\eea
The solution becomes the usual magnetic RN black hole when $\gamma=0$.
For $\gamma <0$, the curvature singularity is located at $r=0$ and hence there must be a horizon $r=r_0>0$, where $r_0$ is the largest root of $f$.  If $\gamma>0$, there is an additional curvature singularity located at $r_*=(2\gamma p^2)^{\fft14}$, and we must require that $r_0>r_*$. This implies
\be
\mu>\fft{\pi(3-32\gamma\Lambda_0)p^{\fft32}}{32*2^{\fft34} \gamma^{\fft14}} +
\fft{3\epsilon\gamma^{\fft14} \sqrt{p}\, \Gamma(\fft34)^2}{2^{\fft34}\sqrt{\pi}}\,.
\ee
Once the event horizon $r_0$ exists, then the temperature and entropy are given by
\bea
T=\fft{-p^2 + 4\epsilon r_0^2 - 4\Lambda_0 r_0^2}{16\pi r_0^2 (r_0^4 - 2\gamma p^2)^{\fft14}}\,,\qquad S=\pi r_0^2 \Big(1 + \fft{2\gamma p^2}{r_0^4}\Big)\,.
\eea
The solution becomes extremal with $T=0$ if
\be
p^2=4\epsilon r_0^2 -4\Lambda_0 r_0^2\,.
\ee
The mass and magnetic charge of the black hole are given by
\be
M=\ft12\mu\,,\qquad Q_m=\ft14 p\,.
\ee
We do not have an independent way of determining the thermodynamical potential $\Phi_m$ for the magnetic charge, and we determine it by completing the first law of the black hole thermodynamics
\be
dM = T dS + \Phi_m dQ_m\,.\label{magfirstlaw}
\ee
We find a complicated expression
\bea
\Phi_m &=&\fft{p \big(7 r_0^4 + 2 \gamma(p^2 - 32\epsilon r_0^2 + 32\Lambda_0 r_0^4)\big)}{
16r_0^4(r_0^4 - 2\gamma p^2)^{\fft14}} +\fft{2\epsilon\gamma\,p}{r_0^3}\,
{}_2F_1[\ft14,\ft34;\ft74;\ft{2\gamma p^2}{r_0^4}]\cr
&&+\fft{3p}{16r_0}(3-32\gamma\Lambda_0)\, {}_2F_1[\ft14,\ft14;\ft54;
\ft{2\gamma p^2}{r_0^4}]\,.
\eea
Although we determine the $\Phi_m$ using the first law (\ref{magfirstlaw}), the result is nontrivial since the first law (\ref{magfirstlaw}) involves two independent parameters and hence a non-trivial integrability condition. To be specific, it is non-trivial in our case that $(dM-TdS)$ does not involve terms proportional to $dr_0$, which would make the first law invalid.

\subsection{Exact electric $z=2$ Lifshitz black holes}

When $p=0$, the ansatz carries only the electric charge $Q_e=\fft14q$.  We have not find the general exact solutions for generic parameters.  However, when
\be
8\gamma \Lambda_0=3\,,
\ee
we find an exact solution for general $q$:
\bea
f=g^2 r^2 + \epsilon - \fft{g q \sqrt{r^2 + 4\mu}}{2r}\,,\qquad
h=(r^2 + 4\mu) f\,,\qquad\phi=g (r^2-r_0^2)\,,
\eea
where the constant $g$ is defined by $\Lambda_0=-3g^2$ and $r_0$ is the location of the event horizon defined by $f(r_0)=0$.  When $\mu=0=q$, the solution describes
a Lifshitz vacuum of $z=2$, namely
\be
ds^2=-r^2(g^2r^2 +\epsilon) dt^2 + \fft{dr^2}{g^2 r^2 + \epsilon} + r^2 \Omega_{2,\epsilon}^2\,.
\ee
(To be precise, the Lifshitz metric is given by $\epsilon=0$, in which case the spacetime is homogeneous.  For non-vanishing $\epsilon$, the metric has curvature singularity at $r=0$.)  The large-$r$ expansion of $f$ is given by
\be
f=g^2 r^2 + \epsilon - \ft12 g q - \fft{g q \mu}{r^2} + \cdots\,.
\ee
It follows from \cite{Liu:2014dva} that the mass can be read off as $M=\ft12 gq\mu$. The first law of thermodynamics
\be
dM=TdS + \Phi_e dQ_e\,,
\ee
can be easily verified where the thermodynamics quantities are
\bea
T&=&\fft{\sqrt{f'(r_0) h'(r_0)}}{4\pi}\,,\qquad S=\pi r_0^2\,,\cr
M&=& \ft12 g q \mu\,,\qquad Q_e=\ft14 q\,,\qquad \Phi_e=-g(r_0^2 + 2\mu)\,.
\eea
For $\epsilon=0$, it has generalized Smarr relation $M=\ft12 (TS +\Phi_eQ_e)$.  Note that this is different from the generalized Smarr relation for the AdS planar black holes (\ref{smarr1}).

\subsection{Dyonic black holes}

In four dimensions, the Maxwell field in a black hole can carry both electric and magnetic charges, giving rise to dyonic solutions. we do not have exact solutions for such general parameters.  We find two approximate solutions, one for small $\gamma$ and the other for small charges $(p,q)$.

\subsubsection{Small-$\gamma$ black holes}

We first present the small $\gamma$ solutions. When $\gamma=0$, the solutions are the dyonic RN black holes.  At the linear order of $\gamma$, we find
\bea
f &=&\bar f (1 + \gamma \tilde f) + {\cal O}(\gamma^2)\,,\qquad \bar f=g^2 r^2 +\epsilon - \fft{\bar \mu}{r} + \fft{q^2 + p^2}{4r^2}\,,\cr
u &=& 1 + \fft{\gamma (q^2 - 3p^2)}{r^4} + {\cal O}(\gamma^2)\,,\cr
\phi &=& \phi_0 - \fft{q}{r_0} + \fft{\gamma (7p^2 q^2 + 3 q^3 - 20\mu q r + 80 g^2 q r^4}{10 r^5} + {\cal O}(\gamma^2)\,,
\eea
where
\be
\tilde f= \fft{7p^2-q^2}{2r^4} +\Big(\fft{c_1}{4r} - \fft{(p^2 + q^2)(3p^2-q^2 - 20\epsilon r^2)}{40r^6} + \fft{3g^2 (3p^2-q^2)}{2r^2}\Big)\fft{1}{\bar f}
\ee
For the small-$\gamma$ approximation to be valid for all regions on and out of the horizon, $\tilde f$ must be well-defined for $r\ge r_0$ where $\bar f(r_0)=0$.  This condition restricts the parameter $c_1$, namely
\be
c_1 = \fft{(p^2 + q^2)(3p^2-q^2 - 20\epsilon r_0^2)}{10r_0^5} - \fft{6g^2 (3p^2-q^2)}{r_0}\,.
\ee
Now the solution describes a dyonic black hole for sufficiently small $\gamma$.  The asymptotic large-$r$ expansion of the function $f$ is give by
\be
f = g^2r^2 + \epsilon - \fft{\mu}{r} + \fft{p^2 + q^2 + 8g^2\gamma (4p^2-q^2)}{4r^2} + \cdots\,,
\ee
where $\mu=\bar\mu -\ft14 c_1 \gamma$.  Thus the mass and electric and magnetic charges are
\be
M=\ft12\mu\,,\qquad Q_e=\ft14 q\,,\qquad Q_m=\ft14 p\,.
\ee
The other thermodynamic quantities, up to the linear $\gamma$ order, are given by
\bea
T &=& \fft{4r_0^2(3g^2 r_0^2 -\epsilon)-p^2-q^2}{16\pi r_0^2} +\fft{\gamma\big((12g^2r_0^4-p^2+q^2)(p^2+q^2) + 4\epsilon (p^2-3q^2)\big)}{32\pi r_0^7},\cr
S &=& \pi r_0^2 \big(1 + \fft{2\gamma p^2}{r_0^4}\big)\,,\qquad
\Phi_e= \fft{q}{r_0} -\fft{\gamma q (7p^2 + 3 q^2 -20\bar \mu r_0 + 80 g^2 r_0^4)}{10r_0^6}\,,\cr
\Phi_m &=& \fft{p}{r_0} + \fft{2\gamma p \big(5 (3g^2 r_0^2 -\epsilon)r_0^2 + p^2 + 2q^2\big)}{5r_0^5}\,.
\eea
It is then straightforward to verify that the first law of black hole thermodynamics
\be
dM=TdS + \Phi_e dQ_e + \Phi_m dQ_m\,,\label{dyonfirstlaw}
\ee
is valid up to and including the linear order of $\gamma$.

The purely electric small-$\gamma$ solution ($p=0$) was obtained in \cite{Cai:2011uh}, where thermodynamical properties were analysed using Euclidean action approach based on the quantum statistic relation (QSR)\cite{Gibbons:1976ue}. Our results disagree with this approach.  Such a phenomenon also occurred in Einstein-Horndeski gravity and it was suggested that the culprit is that the theory may not have a Hamiltonian formalism \cite{Feng:2015oea,Feng:2015wvb}.  We expect that the same situation occurs here.  Our example serves a further lesson that the QSR becomes problematic in theories with non-minimally coupled derivative matter fields.

\subsubsection{Small charge black holes}

An alternative approximation is to consider small charges.  The leading-order solution is then the Schwarzschild black hole with
\be
\bar f=g^2 r^2 + \epsilon -\fft{\bar \mu}{r}\,.
\ee
We find that up to and including the quadratic order of electric and magnetic charges, the solutions are
\bea
f &=& f_0 (1 + \tilde f)\,,\qquad u=1 + \tilde u\,,\qquad \phi=\tilde \phi\,,\cr
\tilde f &=& -\fft{1}{r \bar f} \Big(c_1 + \ft14 Q(r) +
\fft{p^2(2\gamma (7\bar \mu - 8\epsilon r - 16g^2 r^3)-r^3)}{4r^3}\Big)\,,\cr
\tilde u &=& 1 + \fft{Q(r)}{6\bar\mu} -\fft{3\gamma p^2}{r^4} +
\fft{q^2 r^2}{6\bar \mu ((8g^2\gamma+1) r^3 - 8\gamma \bar \mu)}\,,\cr
\tilde\phi &=& \fft{Q(r)}{q}\,.
\eea
where
\bea
Q(r) &=& \int_\infty^r \fft{q^2 r'}{(8g^2\gamma +1) r'^3 - 8\gamma \bar\mu}\, dr'\cr
&=&\fft{q^2}{12\sqrt3(\gamma\bar\mu (8\gamma g^2 +1)^2)^{\fft13}}\Big[-6\arctan
\Big(\fft{\sqrt3}{1 + (\fft{8\gamma g^2 +1}{\gamma\mu})^{\fft13} r}\Big)\cr
&&+ 3\sqrt3 \log \big((1 + 8\gamma g^2)^{\fft13} - 2(\gamma\bar\mu)^{\fft13}\big) -
\sqrt3 \log \big((8\gamma g^2+1)r^3 - 8\gamma \bar \mu\big)\Big]\,.
\eea
For the expansion to be valid, the horizon $r=r_0$ with $\bar f(r_0)=0$ should not be altered.  This implies that
\be
c_1 = - \ft14 Q(r_0) - \fft{p^2(2\gamma (7\bar \mu - 8\epsilon r_0 - 16g^2 r_0^3)-r_0^3)}{4r_0^3}\,.
\ee
The thermodynamical quantities can now be easily calculated, given by
\bea
&&M=\ft12\mu = \ft12(\bar \mu + c_1)\,,\qquad  T=\fft{f'(r_0)\sqrt{u(r_0)}}{4\pi}\,,\qquad S=\pi r_0^2 \big(1 + \fft{2\gamma p^2}{r_0^4}\big)\,,\cr
&&Q_e=\ft14 q\,,\qquad \Phi_e=-\fft{Q(r_0)}{q}\,,\cr
&&Q_m=\ft14 p\,,\qquad \Phi_m=\fft{p((6\gamma g^2 +1) r_0^2 - 2\epsilon\gamma)}{r_0^3}\,.
\eea
It is now straightforward to verify that the first law (\ref{dyonfirstlaw}) is indeed satisfied up to and including the quadratic order of the electric and magnetic charges.

\section{AdS/CFT application: viscosity/entropy ratio}

Having constructed theory and obtained many charged black hole solutions, we are in the position to discuss applications in the AdS/CFT correspondence.  One such an application is that the AdS planar or Lifshitz black holes are dual to some ideal fluid and the linear response of a graviton in the $SO(2)$-rotational invariant directions can be used to calculate the shear viscosity of the fluid \cite{KSS,KSS0}.  In two-derivative gravities, various arguments were given that the viscosity/entropy ratio is fixed, given by
\be
\fft{\eta}{S}=\fft{1}{4\pi}\,.
\ee
This value is no longer held in higher-derivative gravities \cite{shenker}. There is no universal answer, which depends on the details of theories such as coupling constants, as well as the integration constants of the solution such as mass and charges.

For higher-derivative gravities, there is typically a shortcoming in literature that the results are applicable only for small coupling constants of the higher-derivative terms
\cite{Cai:2011uh,Myers:2010pk,Myers:2009ij,Ritz:2008kh,Lemos:2011gy}.  This may be the consequence of two obstacles. One is that the higher-derivative theory is only defined for the small couplings, as in the case of perturbative string theory.  The theory would have ghost issue when treated on its own.  This issue is resolved by our construction so that the theory can be ghost free.  Another obstacle is that exact solutions may be lacking for higher-derivative gravities for general parameters.  This is indeed the case for our theory.  Although we have find many exact examples of special solutions, we do not have the exact solutions of the most general dyonic black holes for the generic parameters.

Recently new technique was developed where the viscosity can be calculated without knowing the exact solutions \cite{Liu:2015tqa}.  This technique was developed mainly for two-derivative gravities.  The key point of this technique is that AdS planar black holes or Lifshitz black holes have a scaling symmetry that gives rise a Noether charge which relates the quantity on the horizon to that on the asymptotic infinity.  The consequence is a generalized Smarr relation, which can be viewed as the bulk dual to the boundary viscosity/entropy relation. Since the existence of the Noether charge associated with the scaling symmetry is independent of the number of derivatives of the theory,
we find that this technique can be adopted for our higher-derivative gravities as well.  Thus although we do not have the general solutions for equations (\ref{einseom}), the equations themselves are enough for us to determine the viscosity/entropy ratio.

To proceed, we set $\Lambda_0=-3g^2$. It is important to note that we are now dealing with the case $\epsilon=0$. It follows from the equation (\ref{einseom}) that we have
\be
f'(r_0)=\fft{r_0(3g^2 r_0^4 - p^2 -q^2)}{4(r_0^4 - 2\gamma p^2)}\,.
\ee
The temperature is therefore
\be
T=\fft{r_0(3g^2 r_0^4 - p^2 -q^2)\sqrt{u(r_0)}}{16\pi(r_0^4 - 2\gamma p^2)}\,.
\ee
To derive the shear viscosity, we consider the traceless and transverse perturbation on the metric
\be
d\Omega_{2,\epsilon=0}^2 = dx_1^2 + dx_2^2 \rightarrow dx_1^2 + dx_2^2 + 2\Psi(r,t) \,dx_1 dx_2\,.
\ee
The graviton mode $\Psi(r,t)$ satisfies
\be
\ddot \Psi - hf\, \Psi'' - \fft{h \big( r (hf)' -4 hf\big) + 2\gamma f\phi'
\big( (4hf \phi')' - r(hf' - 5 f h') \phi'\big)}{2r (h-2\gamma f\phi'^2)} \,\Psi'=0\,,\label{waveeom}
\ee
together with the constraint
\be
\gamma\, p\,\big( \fft{f}{h} \phi'^2\big)'\,\dot \Psi=0\,.\label{cons}
\ee
The constraint arises in the linearized Einstein equations in the diagonal $(x_1,x_1)$ and $(x_2,x_2)$ directions, whilst the wave equation (\ref{waveeom}) arises in the off-diagonal
$(x_1,x_2)$ direction.  The constraint (\ref{cons}) is automatically satisfied for general dyonic black holes in the Einstein-Maxwell theory, corresponding to $\gamma=0$. For non-vanishing $\gamma$, the constraint is satisfied only for either purely electric solution or purely magnetic solution, but not for the general dyonic solution, i.e. we need to impose
\be
Q_e Q_m=0\,.\label{nodyoncond}
\ee
It turns out that the wave equation (\ref{waveeom}) can be analysed without imposing the condition (\ref{nodyoncond}), and hence we shall thus proceed.  Making a Fourier transformation in time
\be
\Psi(r,t) = e^{-{\rm i}\omega t} \psi(r)\,,
\ee
we find, near the horizon, that $\psi$ satisfies
\be
(r-r_0)^2 \psi'' + (r-r_0)\psi' + \fft{\omega^2}{16\pi^2 T^2}\, \psi=0\,.
\ee
This equation can be solved exactly, implying
\be
\psi =\psi_0 e^{-\fft{{\rm i}\omega}{4\pi T}\log(r-r_0)} \sim e^{-\fft{{\rm i}\omega}{4\pi T}\log(\fft{f}{g^2r^2})}\Big|_{r\rightarrow r_0}\,.
\ee
In other words, we select only the ingoing modes. To extend the horizon solution to asymptotic infinity, we make the following ansatz
\be
\psi =\psi_0 e^{-\fft{{\rm i}\omega}{4\pi T}\log(r-r_0)} \sim e^{-\fft{{\rm i}\omega}{4\pi T}\log(\fft{f}{g^2r^2})}\Big(1 - {\rm i}\omega\, U(r) +{\cal O}(\omega^2)\Big)\,.
\ee
where $U$ should be regular on the horizon and vanish at the asymptotic infinity.
At the linear order of $\omega$, we find that the function $U$ is a quadrature, given by
\bea
U' &=&\fft{V}{r^2(h + 2\gamma f \phi'^2)}\,\sqrt{\fft{h}{f}}\,,\cr
V &=& V_0 -\fft{r\big(r^4 + 2\gamma (8r^2 f+q^2) + 64\gamma^2 f^2\big)}{16\pi T (r^4-2\gamma p^2)(r^2 + 8\gamma f)^3}\Big( - q^2 r^4 \cr
&&\qquad\qquad+(r^2 + 8\gamma f)\big(12 r^4(g^2r^2 -f)-
p^2(r^2 + 32 \gamma f)\big)\Big)\sqrt{u}\,.
\eea
In order for $U$ to be regular on the horizon, we must have $V(r_0)=0$, which implies
\be
V_0=\fft{(r_0^4-2\gamma q^2)(12g^2 r_0^4-p^2 -q^2)\sqrt{u(r_0)}}{16\pi T\,r_0
(r_0^4 - 2\gamma p^2)}=r_0^2\Big(1 + \fft{2\gamma q^2}{r_0^4}\Big)\,.
\ee
To extract the information of the shear viscosity of the boundary field theory, we consider the effective Lagrangian for $\psi$, given by
\be
{\cal L} \sim \fft{1}{16\pi} r^2 \sqrt{\fft{f}{h}} ( h + 2\gamma f\phi'^2)\, \psi'^2 + \cdots\,,
\ee
Thus the action can be evaluated, given by
\bea
I &\sim& \fft{1}{16\pi}\int \Omega_{\2}^{\epsilon=0}\, \Big( r^2 \sqrt{\fft{f}{h}} ( h + 2\gamma f\phi'^2)\, \psi' \psi\Big)\Big|_{\Sigma}\cr
&=& -{\rm i} \omega\, \fft{r(h + 2\gamma f\phi'^2)}{16\pi\sqrt{hf}}\Big(
4\pi r f\, U' + \fft{r f'-2f}{T}\Big) + {\cal O}(\omega^2)\,,\cr
&=& -{\rm i}\omega \ft14 V_0 +  {\cal O}(\omega^2)\,.
\eea
The shear viscosity can then be read off
\be
\eta = \ft14 V_0= \ft14 r_0^2 \Big(1 + \fft{32\gamma Q_e^2}{r_0^4}\Big)\,.\label{vis}
\ee
It follows from the definition of the entropy (\ref{entropy}) that the viscosity/entropy ratio is then given by
\be
\fft{\eta}{S} = \fft{1}{4\pi}\, \fft{r_0^4 + 32\gamma Q_e^2}{r_0^4 + 32\gamma Q_m^2}\,.\label{visent}
\ee
Thus we obtain the ratio without using any exact solution. Owing to the constraint (\ref{nodyoncond}), the result is applicable for all purely electric or purely magnetic black holes, for all ranges of $\gamma$ where a black hole exists.

The viscosity (\ref{vis}) was obtained also in \cite{Cai:2011uh} for the small $\gamma$ parameter for which the approximate solution was found.  Our general result confirms this.  However, our viscosity/entropy ratio (\ref{visent}) disagrees with \cite{Cai:2011uh} even for vanishing $Q_m$ and small $\gamma$.  This is because the entropy in \cite{Cai:2011uh} was obtained using the Euclidean action procedure, which we believe is invalid in this theory.

\section{Conclusions}

In this paper we constructed higher-derivative gravities with a non-minimally coupled Maxwell field $A_\1=A_\mu dx^\mu$.  The general Lagrangian consists of invariant polynomials built from the Riemann tensor and the field strength $F_\2=dA_\1$.  These polynomials are analogous to the Euler integrands in Lovelock gravities in that the field equations of motion remains second order for both the metric and $A_\mu$.  The total higher derivatives are achieved through nonlinearity.  The linearized equations of motion in any background involves only two derivatives and hence the theories can be ghost free. We also generalize the construction to involve a generic non-minimally coupled $p$-form field strength. We noted that as $p$ increases, more and more invariant polynomials could be constructed to give rise to ghost-free theories.  However, we did not classify all possible structures.

As an application in black hole physics, we focused on a low-lying example in which the Einstein-Maxwell gravity with a cosmological constant was augmented by a polynomial built from the Riemann tensor and bilinear $F_\2$, with a coupling constant $\gamma$.  We constructed charged static black holes in four dimensions with isometries of $S^2$, $T^2$ and $H^2$.  Although we do not have the most general exact solutions, we obtained many exact special ones, including the magnetic black holes and also electrically charged Lifshitz black holes with critical exponent $z=2$.  We then constructed analytic approximate dyonic solutions with small charges or with small parameter $\gamma$.  We studied the thermodynamics of the black holes and obtained the general first law.  An important lesson is that the first law based on Wald formalism disagrees with that from the Euclidean action procedure based on QSR.  Such phenomenon was first observed in Einstein-Horndeski gravity and it was suspected that Einstein-Horndeski gravity may not admit a Hamiltonian formalism \cite{Feng:2015oea}.  Our results suggest this may be a widespread situation for theories involving non-minimally coupled form fields.

We then studied an application of the theory in the AdS/CFT correspondence by deriving the boundary viscosity/entropy ratio for AdS or Lifshitz planar black holes.  The purpose of our work is that higher-derivative terms in our theory do not have to be small and the theory can stand on its own right.  The lacking of the exact general solution appears to produce an obstacle to get general results for all allowed parameters.  We find that the viscosity/entropy ratio can be fully determined without needing to know the black hole solutions; the equations of motion suffice.  We thus obtain the viscosity/entropy ratio for all parameters, including the coupling constant $\gamma$ and electric and magnetic charges, none of which has to be small.

Form fields arise naturally in string and M-theory.  They typically couple to gravity non-minimally in higher-order expansions of the low-energy effective theories of the purturbative strings.  Our ghost-free construction makes it possible to treat the theories in finite order and study the theories on their own right. The explicit results of black holes and their certain AdS/CFT application in the low-lying example shows rich structures that deserve further investigation.

\section*{Acknowledgement}

The work is supported in part by NSFC grants NO. 11475024 and NO. 11235003.

\end{document}